\documentclass{PoS}

\usepackage{amsmath}
\usepackage{graphicx}

\title{Convergent Perturbation Theory for the lattice $\phi^4$-model}

\ShortTitle{CPT for the lattice $\phi^4$-model}

\author{Vladimir V. Belokurov\\
        Lomonosov Moscow State University, Russia,\\
        Institute for Nuclear Research RAS, Russia\\
        E-mail: \email{vvbelokurov@yandex.ru}}
\author{\speaker{Aleksandr S. Ivanov}\\
        Lomonosov Moscow State University, Russia,\\
        Institute for Nuclear Research RAS, Russia\\
        E-mail: \email{ivanov.as@physics.msu.ru}}
\author{Vasily K. Sazonov\\
        University of Graz, Austria\\
        E-mail: \email{vasily.sazonov@uni-graz.at}}
\author{Eugeny T. Shavgulidze\\
        Lomonosov Moscow State University, Russia\\
        E-mail: \email{shavgulidze@bk.ru}}

\abstract{
The standard lattice perturbation theory leads to the asymptotic series because of the incorrect 
interchange of the summation and integration. However, changing the initial approximation of the perturbation theory, 
one can generate the convergent series. We study the lattice $\phi^4$-model and compare the operator $\langle\phi_n^2\rangle$ 
calculated using the convergent series and obtained by Monte Carlo simulations.  
}

\FullConference{The 33rd International Symposium on Lattice Field Theory\\
		14 -18 July 2015\\
		Kobe International Conference Center, Kobe, Japan*}

\begin{document}

\section{Introduction}
Typically the small coupling expansions in quantum field theories and lattice models have an asymptotic character
and do not converge \cite{Dyson, Lipatov}. The reason for this is hidden in the incorrect interchanging of the summation and integration.
Nevertheless, considering a suitable regularization of the path or lattice integrals 
\cite{Belokurov1, Belokurov2, Belokurov3, BelSolSha97, BelSolSha99, Meurice2004, Meurice20052} or changing
the initial Gaussian approximation to an appropriate interacting theory \cite{Shaverdyan1983, UshveridzeSuper}
one can construct convergent series (CS). Recently, the applications of the convergent series methods
to a model of lattice QED \cite{Sazonov2014} and to the continuous Yang-Mills and QCD \cite{Sazonov2015} were proposed.
This gives a strong motivation for the detailed studies of the CS. 
In the current work we test the approach \cite{Shaverdyan1983, UshveridzeSuper}
applied to a lattice model, leaving the methods \cite{Belokurov1, Belokurov2, Belokurov3, BelSolSha97, BelSolSha99, Meurice2004, Meurice20052}
for the future investigations. Initially, the method \cite{Shaverdyan1983, UshveridzeSuper} 
was developed for the scalar field theories
using the assumption of the applicability of the dimensional
regularization. Such assumption is not needed in case of the lattice models
and the construction of the convergent series can be done rigorously.
At the same time the results obtained using the CS for the lattice models can be directly compared with the Monte Carlo
simulations. Here we present the study of the one dimensional lattice $\phi^4$-model 
with the periodic boundary conditions within the framework of the convergent series.
The model is determined by the action
\begin{equation}
S[\phi_n] = \frac{1}{2} \sum_{m,n = 0}^N \phi_m K_{mn} \phi_n + \frac{\lambda}{4!} \sum_{n = 0} ^ N \phi _ {n} ^ 4,
\label{Sphi4}
\end{equation}
where
\begin{equation}
\frac{1}{2} \sum_{m,n = 0}^N \phi_m K_{mn} \phi_n = 
\sum_{n = 0} ^ N \Big[ -\frac{1}{2} \phi _ {n} \phi _ {n + 1} - \frac{1}{2} \phi _ {n} \phi _ {n - 1} + \phi _ {n} ^ 2 + \frac{1}{2} M ^ 2 \phi _ {n} ^ 2 \Big]
\end{equation}
and the subject of our computations is the normalized to the full partition function operator
\begin{equation}
\langle\phi_n^2\rangle = \frac{\int\, \prod_{x}\, [d\phi_x]\, \phi_n^2\, e^{-S}}{\int \, \prod_{x}\, [d\phi_x]\, e^{-S}}\,,
\end{equation}
which is the propagator $\langle\phi_i \phi_j\rangle$ with the coinciding $i$ and $j$.

\section{Description of the CS-method}
Let us consider the propagator $\langle\phi_i \phi_j\rangle$ normalized to the free theory 
\begin{eqnarray}
  \langle\phi_i \phi_j\rangle = \frac{1}{Z_0}\prod_{n}^V \int\, [d\phi_n]\, \phi_i \phi_j\, \exp\{-S[\phi_n]\}\,,
\label{correl}
\end{eqnarray}
where
\begin{eqnarray}
  Z_0 = \prod_{n}^V \int\, [d\phi_n]\, e^{-\frac{1}{2} \phi_m K_{mn} \phi_n}
\end{eqnarray}
is the partition function of the free theory and $V$ is the volume of the lattice.
Adapting the approach from \cite{Shaverdyan1983, UshveridzeSuper} for the lattice, 
we construct convergent series splitting the action into the
new non-perturbed part $N[\phi]$ and perturbed part $P[\phi]$ as
\begin{equation}
  S[\phi_n] = N[\phi_n] + P[\phi_n] = N[\phi_n] + (S[\phi_n] - N[\phi_n])
\end{equation}
with
\begin{equation}
  N[\phi_n] = \sum_{n, m} \frac{1}{2} \phi_n K_{n m} \phi_m + \sigma \Big(\sum_{n, m} \frac{1}{2} \phi_n K_{n m} \phi_m\Big)^2\,.
\end{equation}
Then it is possible to show \cite{Ivanov2015} that for $\sigma > \frac{\lambda}{6 M^4}$ follows that $N[\phi_n] \geq S[\phi_n]$
and the propagator can be obtained as a sum of the convergent series
\begin{equation}
  \langle\phi_i \phi_j\rangle = \frac{1}{Z_0} \sum_{l = 0}^\infty \langle\phi_i \phi_j\rangle_l
\label{sum}
\end{equation}
with terms given by
\begin{equation}
  \langle\phi_i \phi_j\rangle_l = \frac{1}{l!} \prod_{n}^V \int\, [d\phi_n]\, \phi_i \phi_j\, \{N[\phi_n] - S[\phi_n]\}^l\, \exp\{-N[\phi_n]\}\,.
\label{lf}
\end{equation}
The functions (\ref{lf}) can be calculated in the following way. Introducing an
auxiliary integration we change $\lVert\phi_n\rVert \equiv \Big(\frac{1}{2} \phi_n K_{n m} \phi_m\Big)^{\frac{1}{2}}$
to the one dimensional variable $t$
\begin{equation}
  \langle\phi_i \phi_j\rangle_l = \frac{1}{l!} \int_{0}^\infty dt\, \exp\Big(-t^2 - \sigma t^4 \Big) \prod_{n}^V \int\, [d\phi_n]\, 
  \phi_i \phi_j\,\delta(t - \lVert\phi_n\rVert) 
\Big(\sigma t^4 - \frac{\lambda}{4!} \sum_n \phi_n^4 \Big)^l\,.
\end{equation}
Rescaling the field variables as $\phi_n^{old} = t \phi_n$, we get
\begin{eqnarray}
  \langle\phi_i \phi_j\rangle_l = \frac{1}{l!} \int_{0}^\infty dt\, t^{V + 2} \exp\Big(-t^2 - \sigma t^4\Big) 
  \prod_{n}^V \int\, [d\phi_n]\, \phi_i \phi_j\,\delta(t - t \lVert\phi_n\rVert) \Big(\sigma t^4 - t^4 \frac{\lambda}{4!} \sum_n \phi_n^4 \Big)^l\,,
\label{t1}
\end{eqnarray}
where $t^V$ appeared from
\begin{equation}
  \prod_{n}^V \int\, [d\phi_n^{old}] = t^V \prod_{n}^V \int\, [d\phi_n]\,.
\end{equation}
Employing the identity
\begin{equation}
  \delta(\alpha x) = \frac{\delta(x)}{|\alpha|}\,,
\end{equation}
we rewrite (\ref{t1}) as
\begin{eqnarray}
\nonumber
  \langle\phi_i \phi_j\rangle_l = \frac{1}{l!} \int_{0}^\infty dt\, t^{V + 4 l + 1} \exp\Big(-t^2 - \sigma t^4\Big) \times \\
  \prod_{n}^V \int\, [d\phi_n]\, \phi_i \phi_j\,\delta(1 - \lVert\phi_n\rVert) \sum_{k = 0}^{l} C_l^k \sigma^{l - k} \Big(-\frac{\lambda}{4!} \sum_n \phi_n^4 \Big)^k\,.
\label{fle}
\end{eqnarray}
Now the multi-dimensional (lattice) part of the integral is separated from the auxiliary integration, but it is still not an answer.
The main difficulty is in the calculation of the lattice integral with delta function $\delta(1 - \lVert\phi_n\rVert)$. To solve
this problem we use the following equality
\begin{eqnarray}
  \prod_{n}^V \int\, [d\phi_n]\, \phi_n(x_1)...\phi_n(x_Q)\,e^{-\lVert\phi_n\rVert^2} = 
  \frac{1}{2}\Gamma\Big(\frac{V + Q}{2}\Big) \prod_{n}^V \int\, [d\phi_n]\, \delta(1 - \lVert\phi_n\rVert) \phi_n(x_1)...\phi_n(x_Q)\,.
\label{ident}
\end{eqnarray}
By substituting (\ref{ident}) in (\ref{fle}), we obtain
\begin{eqnarray}
\nonumber
  \langle\phi_i \phi_j\rangle_l  = \frac{1}{l!} \int_{0}^\infty dt\, t^{V + 4 l + 1} \exp\Big(-t^2 - \sigma t^4\Big) \times \\
  \sum_{k = 0}^{l} C_l^k \sigma^{l - k} \frac{2}{\Gamma\Big(\frac{V + 4 k + 2}{2}\Big)}\prod_{n}^V \int\, [d\phi_n]\, \phi_i \phi_j\,
  e^{-\frac{1}{2} \phi_n K_{n m} \phi_m}
  \Big(-\frac{\lambda}{4!} \sum_n \phi_n^4 \Big)^k\,.
\label{main}
\end{eqnarray}
Therefore, each certain order $n$ of the convergent series is expressed as a linear combination of the first $n$ 
orders of the ordinary perturbation theory
with the coefficients given by the one dimensional analytically calculable $t$-depending integrals.
However, the later fact does not mean that the convergent series looses the non-perturbative contributions
from the non-analytical functions such as $e^{-\frac{1}{\lambda}}$. Being an expansion around non-Gaussian initial
approximation it automatically takes non-pertubative contributions into account in the similar way,
as the function $e^{-\frac{1}{\lambda}}$ for positive $\lambda$ can be reproduced by its Taylor series around $\lambda = 1$.

\section{Numerical results}
In the previous section we have described the construction of the convergent series for the propagator $\langle\phi_i \phi_j\rangle$ 
normalized to the free theory. The Monte Carlo simulations provide the propagator normalized to the full partition function.
Then, to compare the results of two different calculations one has to reduce them to a common denominator. We do it
using the fact that in the standard perturbation theory the propagator normalized to the full partition function
can be obtained from the one normalized to the free theory by throwing away the disconnected Feynman diagrams from the expansion.

\begin{figure}[bh]
\centering{
\includegraphics[width=0.9\textwidth]{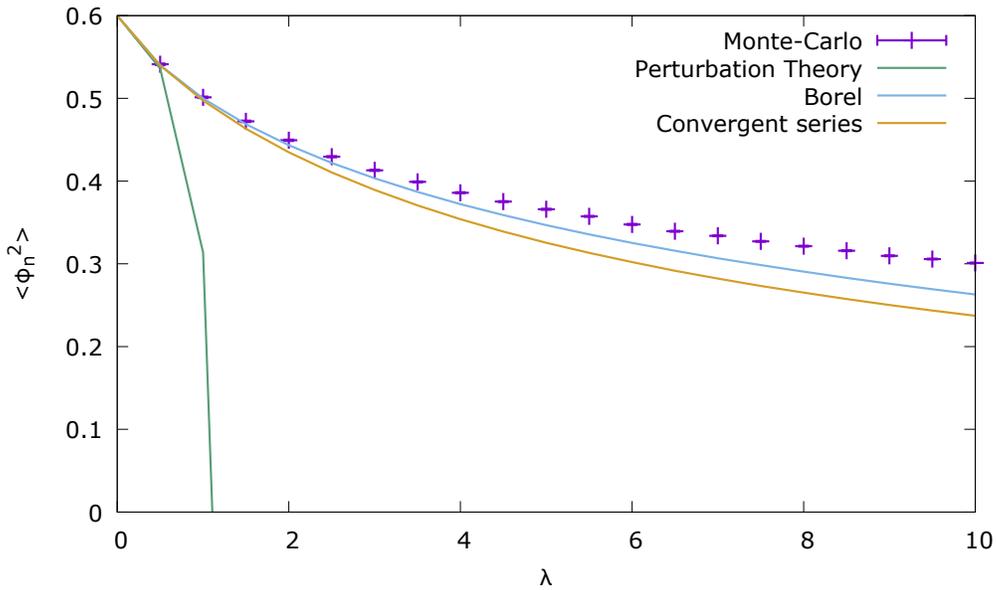}}
\caption{The dependence of $\langle \phi_n^2 \rangle$ on $\lambda$, lattice volume $V = 2$.
The violet dots with the error-bars are the results of the Monte Carlo simulations. 
The green line represents $6$ loops of the standard perturbation theory.
The Borel resummation of $6$ loops of standard perturbation theory is shown by the blue line.
The orange line demonstrates $6$ orders of convergent series.}
\label{N2l010}
\end{figure}

\begin{figure}[bh]
\centering{
\includegraphics[width=0.9\textwidth]{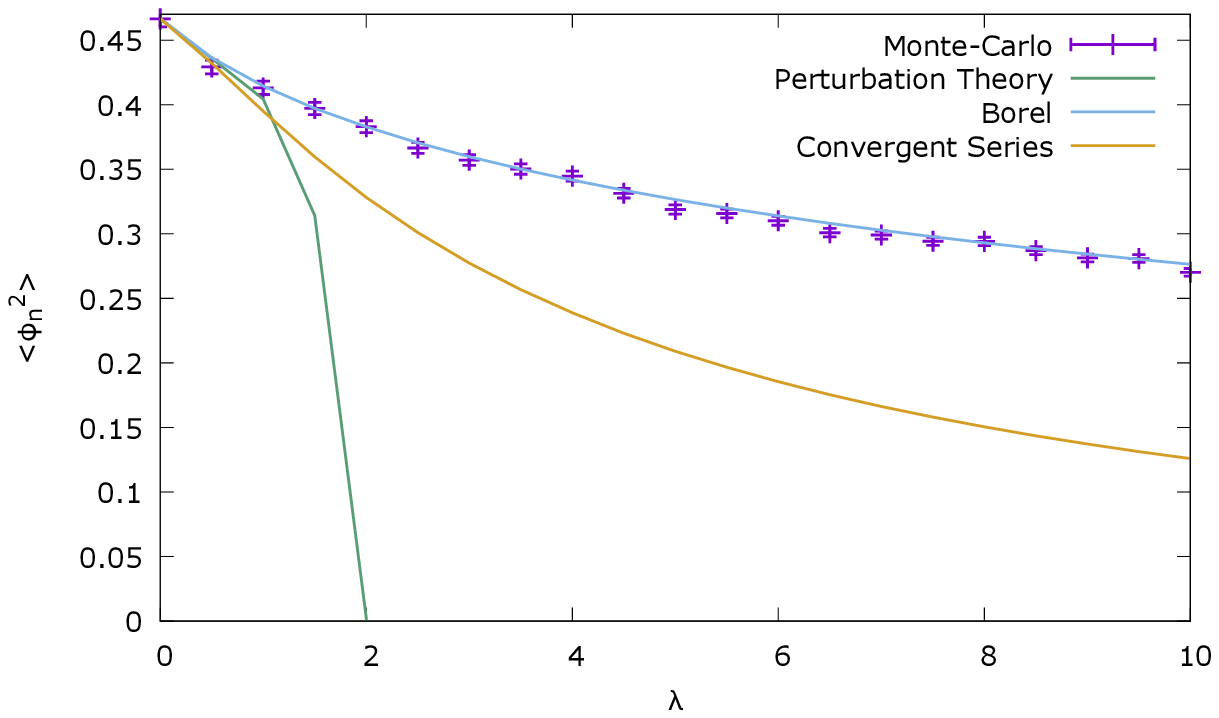}}
\caption{The dependence of $\langle \phi_n^2 \rangle$ on $\lambda$, lattice volume $V = 4$.
The violet dots with the error-bars are the results of the Monte Carlo simulations. 
The green line represents $6$ loops of the standard perturbation theory.
The Borel resummation of $6$ loops of standard perturbation theory is shown by the blue line.
The orange line demonstrates $6$ orders of convergent series.}
\label{N4l010}
\end{figure}

\begin{figure}[bh]
\centering{
\includegraphics[width=0.9\textwidth]{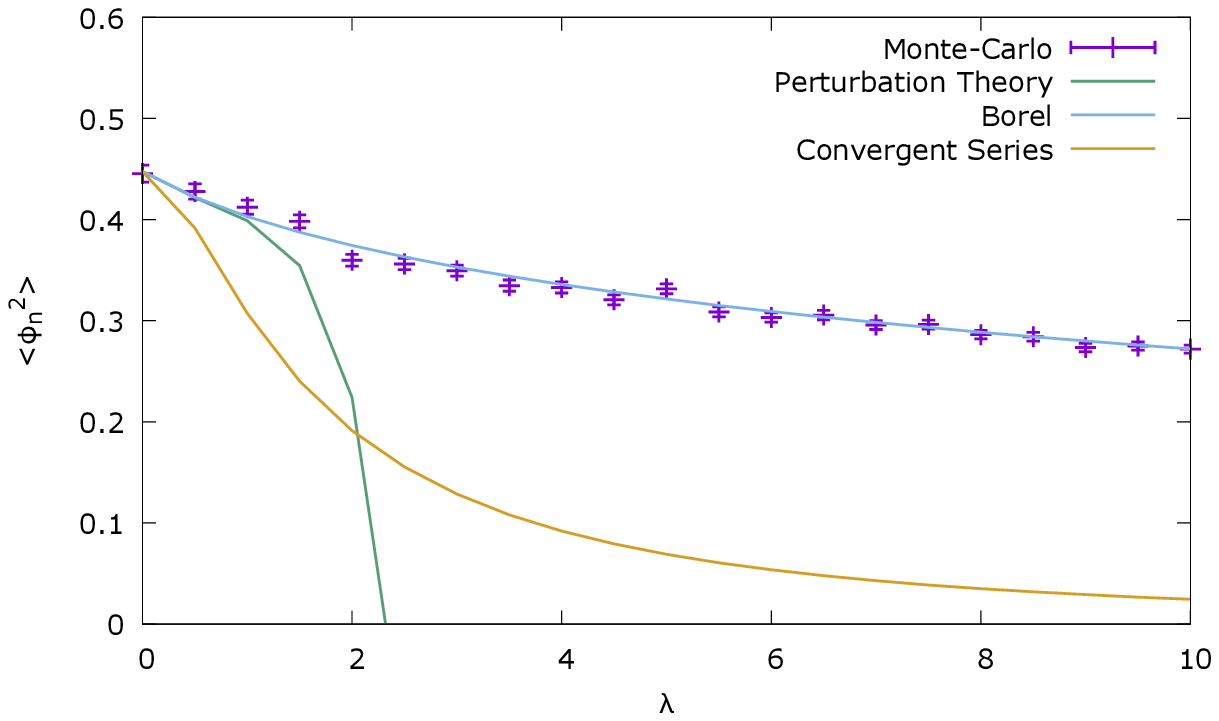}}
\caption{The dependence of $\langle \phi_n^2 \rangle$ on $\lambda$, lattice volume $V = 8$.
The violet dots with the error-bars are the results of the Monte Carlo simulations. 
The green line represents $6$ loops of the standard perturbation theory.
The Borel resummation of $6$ loops of standard perturbation theory is shown by the blue line.
The orange line demonstrates $6$ orders of convergent series.}
\label{N8l010}
\end{figure}

\begin{figure}[bh]
\centering{
\includegraphics[width=0.9\textwidth]{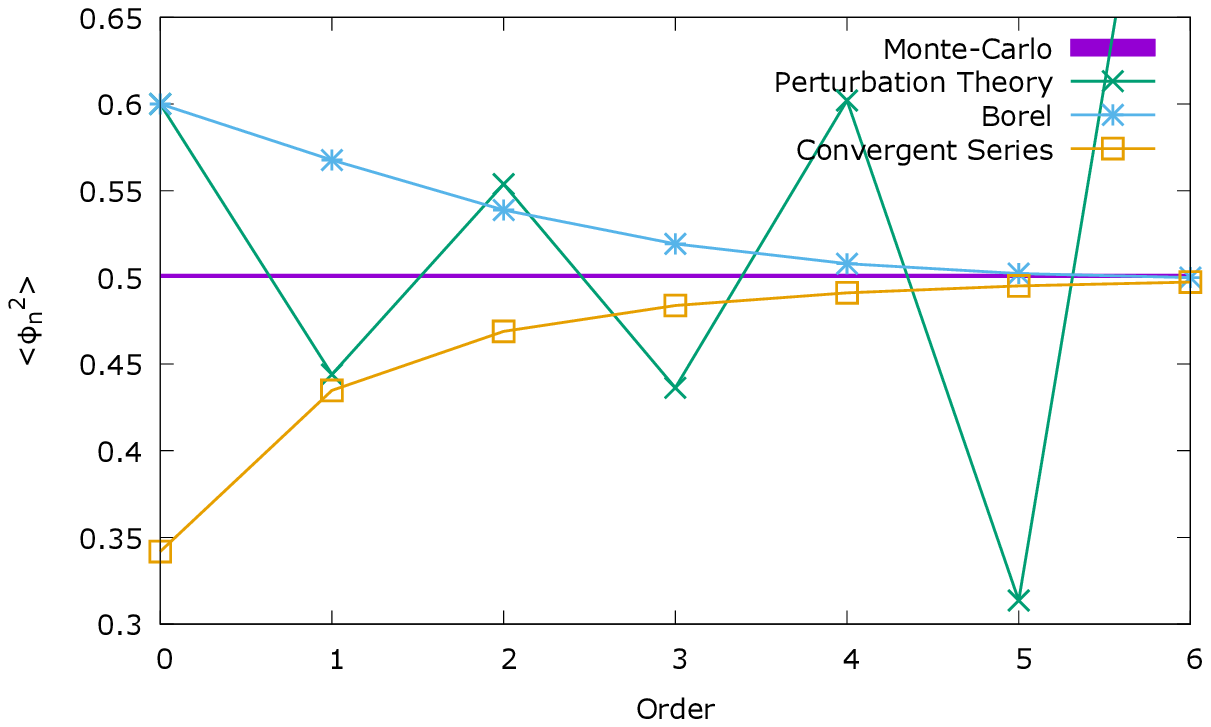}}
\caption{The convergence of the series for $\langle \phi_n^2 \rangle$ at $\lambda = 1$ and lattice volume $V = 2$ 
to the Monte Carlo results depending on the expansion order 
(green line - standard perturbation theory, blue line - Borel resummation,
orange line - convergent series).}
\label{N2l1}
\end{figure}

\begin{figure}[bh]
\centering{
\includegraphics[width=0.9\textwidth]{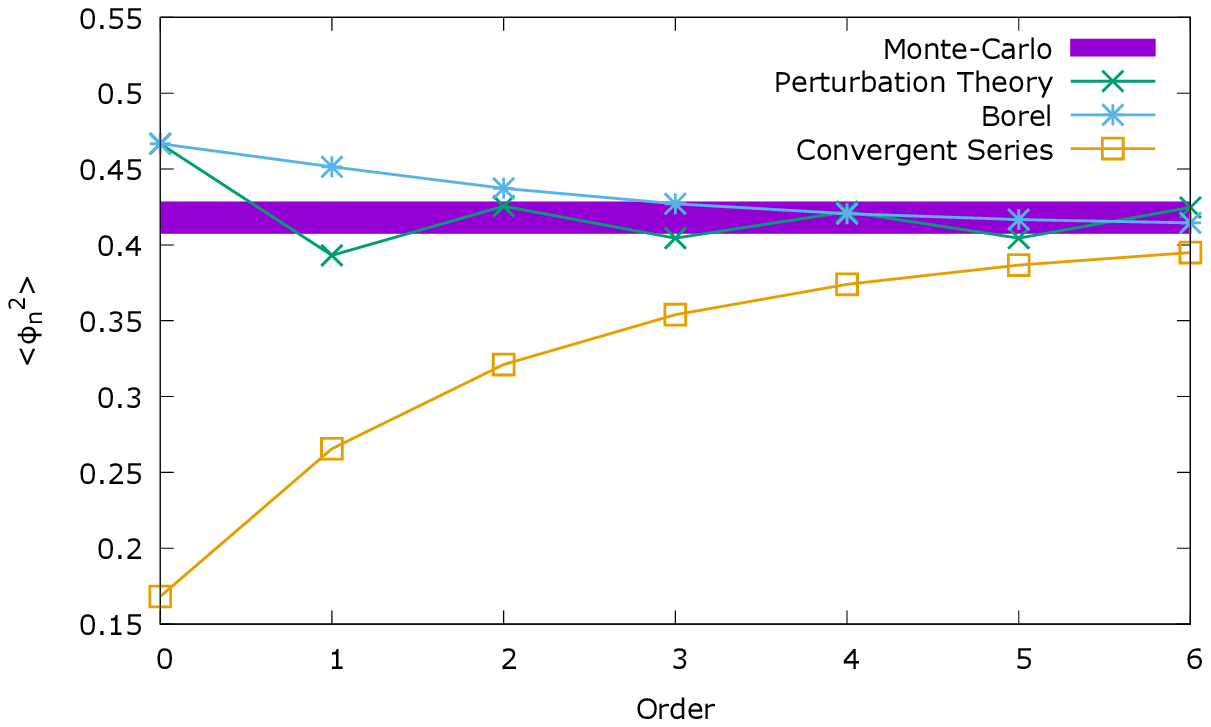}}
\caption{The convergence of the series for $\langle \phi_n^2 \rangle$ at $\lambda = 1$ and lattice volume $V = 4$ 
to the Monte Carlo results depending on the expansion order 
(green line - standard perturbation theory, blue line - Borel resummation,
orange line - convergent series).}
\label{N4l1}
\end{figure}

\begin{figure}[bh]
\centering{
\includegraphics[width=0.9\textwidth]{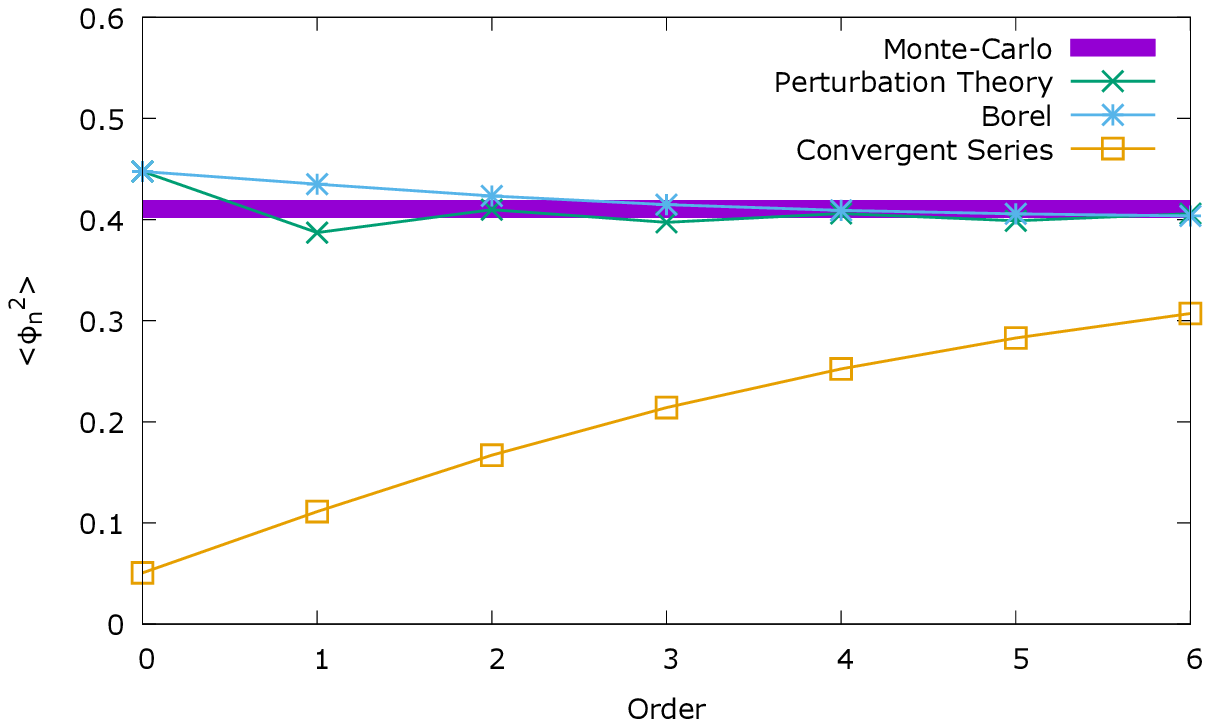}}
\caption{The convergence of the series for $\langle \phi_n^2 \rangle$ at $\lambda = 1$ and lattice volume $V = 8$ 
to the Monte Carlo results depending on the expansion order 
(green line - standard perturbation theory, blue line - Borel resummation,
orange line - convergent series).}
\label{N8l1}
\end{figure}

In Figures \ref{N2l010}, \ref{N4l010} and \ref{N8l010} we present the operator $\langle \phi_n^2 \rangle$ depending on the coupling constant $\lambda$ 
computed within $6$ loops
of the standard perturbation theory,
Borel resummation with conformal mapping and
convergent series in comparison to the Monte Carlo method 
on the lattices with $2$, $4$, and $8$ cites respectively.
\footnote{The Feynman diagrams were generated using the GRACE system \cite{GRACE}.}
In all cases the standard perturbation
theory demonstrate the divergent asymptotic behavior, Borel resummation leads to a good agreement with Monte-Carlo. 
The CS-method agrees with
Monte Carlo for the $2$-site lattice, but deviates from the correct answer more and more for bigger lattice volumes $V$.
The similar picture is seen in Figures \ref{N2l1}, \ref{N4l1} and \ref{N8l1} where we show the convergence of the series to the correct result
depending on the order of the expansion at fixed value of the coupling constant. The cause of the deterioration of matches between CS and Monte Carlo
at large volumes lies in the drastic decrease of the resummation coefficients in (\ref{main}) with growing of 
the lattice volume $V$.

\section{Conclusions}
The studies of the convergent series 
are especially important, because the CS takes into account the non-perturbative contributions 
and, therefore, may provide a complete definition of the path integral.
At the same time, the developing of the convergent series is also important for the lattice computations.
Being constructed from the terms of the standard perturbation theory the CS should not be affected by the sign problem 
\cite{deForcrand} and, therefore, can give its possible solution.
In the present paper we have constructed the convergent series for the one dimensional lattice $\phi^4$-model and computed
the operator $\langle \phi_n^2 \rangle$. For the smallest $2$-site lattice the convergent series method demonstrates a perfect
agreement with Monte Carlo in a wide range of coupling constants. For the larger lattice volumes the convergent series
needs sufficiently more terms to match with the correct results, what makes the direct application of the CS-method 
to the real computations impossible. 
To resolve this problem we have improved the convergent series method.
The derivation and results of the improved CS-method will be presented in our forthcoming paper \cite{Ivanov2015}.
%


\acknowledgments
The work of A.S. Ivanov was supported by the Russian Science
Foundation grant 14-22-00161. V.K. Sazonov acknowledges the  Austrian  Science
Fund FWF, Grant. Nr. I 1452-N27. The analytical derivations were supported by both 
grants and the numerical computations were supported by the Russian Science 
Foundation grant 14-22-00161. 

\clearpage
\label{Bibliography} 
\bibliographystyle{JHEP}
\bibliography{bibliography}

\providecommand{\href}[2]{#2}\begingroup\raggedright\begin{thebibliography}{10}

\bibitem{Dyson}
F.~J. Dyson, {\it Divergence of perturbation theory in quantum
  electrodynamics},  {\em Phys. Rev.} {\bf 85} (Feb, 1952) 631--632.

\bibitem{Lipatov}
L.~L. N. {\em Sov. Phys. JETP} {\bf 45} (1977) 216.

\bibitem{Belokurov1}
V.~Belokurov, Y.~Solov'ev, and E.~Shavgulidze, {\it Method of approximate
  evaluation of path integrals using perturbation theory with convergent
  series. {I}},  {\em Theoretical and Mathematical Physics} {\bf 109} (1996),
  no.~1 1287--1293.

\bibitem{Belokurov2}
V.~Belokurov, Y.~Solov'ev, and E.~Shavgulidze, {\it Method for approximate
  evaluation of path integrals using perturbation theory with convergent
  series. ii. euclidean quantum field theory},  {\em Theoretical and
  Mathematical Physics} {\bf 109} (1996), no.~1 1294--1301.

\bibitem{Belokurov3}
V.~Belokurov, Y.~Solov'ev, and E.~Shavgulidze, {\it Perturbation theory with
  convergent series for functional integrals with respect to the feynman
  measure},  {\em Russian Mathematical Surveys} {\bf 52(2)} (1997) 392.

\bibitem{BelSolSha97}
V.~Belokurov, Y.~Solov'ev, and E.~Shavgulidze, {\it Vychislenie funkcional'nich
  integralov s pomoshyu shodyashehsya ryadov},  {\em Fundament. i prikl.
  matem.} (1997) 693--713.

\bibitem{BelSolSha99}
V.~Belokurov, Y.~Solov'ev, and E.~Shavgulidze, {\it Obshiy podhod k
  vychisleniyu funkcional'nich integralov i summirovaniyu rashodyashehsya
  ryadov},  {\em Fundament. i prikl. matem.} (1999) 363--383.

\bibitem{Meurice2004}
B.~Kessler, L.~Li, and Y.~Meurice, {\it New optimization methods for converging
  perturbative series with a field cutoff},  {\em Phys. Rev. D} {\bf 69} (Feb,
  2004) 045014.

\bibitem{Meurice20052}
L.~Li and Y.~Meurice, {\it A tractable example of perturbation theory with a
  field cutoff: the anharmonic oscillator},  {\em J.Phys. A} {\bf 38} (2005)
  8139--8154, [\href{http://arxiv.org/abs/hep-th/0506038}{{\tt
  hep-th/0506038}}].

\bibitem{Shaverdyan1983}
B.~Shaverdyan and A.~Ushveridze, {\it Convergent perturbation theory for the
  scalar $\phi^{2p}$ field theories; the gell-mann-low function},  {\em Physics
  Letters B} {\bf 123} (1983), no.~5 316 -- 318.

\bibitem{UshveridzeSuper}
A.~Ushveridze, {\it Superconvergent perturbation theory for euclidean scalar
  field theories},  {\em Physics Letters B} {\bf 142} (1984), no.~5-6 403 --
  406.

\bibitem{Sazonov2014}
V.~K. Sazonov, {\it Convergent perturbation theory for lattice {QED}},
  \href{http://arxiv.org/abs/1405.7542}{{\tt arXiv:1405.7542}}.

\bibitem{Sazonov2015}
V.~K. Sazonov, {\it Convergent series for {QCD}},
  \href{http://arxiv.org/abs/1503.00739}{{\tt arXiv:1503.00739}}.

\bibitem{Ivanov2015}
V.~Belokurov, A.~Ivanov, V.~K. Sazonov, and E.~Shavgulidze, {\it In
  preparation.}, .

\bibitem{GRACE}
M.-T. collaboration, {\it Grace system, http://minami-home.kek.jp/}, .

\bibitem{deForcrand}
P.~de~Forcrand, {\it {Simulating {QCD} at finite density}},  {\em PoS} {\bf
  LAT2009} (2009) 010, [\href{http://arxiv.org/abs/1005.0539}{{\tt
  arXiv:1005.0539}}].

\end{thebibliography}\endgroup

%

\end{document}